\documentclass[conference]{IEEEtran}
\IEEEoverridecommandlockouts
\usepackage{cite}
\usepackage{amsmath,amssymb,amsfonts}
\usepackage{algorithmic}
\usepackage{graphicx}
\usepackage{textcomp}
\usepackage{xcolor}
\usepackage{booktabs}
\usepackage{stmaryrd}
\usepackage{comment}
\usepackage{tabularx}
\usepackage{url}
\usepackage{colortbl}
\definecolor{lightgray}{rgb}{0.6, 0.8, 1}
\usepackage{tikz}
\usepackage{adjustbox}

\usepackage{enumerate}

\def\BibTeX{{\rm B\kern-.05em{\sc i\kern-.025em b}\kern-.08em
    T\kern-.1667em\lower.7ex\hbox{E}\kern-.125emX}}
\begin{document}

\title{HW/SW Implementation of MiRitH on\\Embedded Platforms\\
\thanks{This paper was partly funded by the German Federal Ministry of Education and Research as part of the ``PoQ-KIKI'' project under grant number 16KIS2064.\\}
}

\author{
\IEEEauthorblockN{Maximilian Schöffel, Hiandra Tomasi and Norbert Wehn}
\IEEEauthorblockA{\textit{Microelectronic Systems Design Research Group} \\
\textit{RPTU Kaiserslautern-Landau}\\
Kaiserslautern, Germany \\
\{m.schoeffel,tomasi,norbert.wehn\}@rptu.de} 
}

\newcommand\copyrighttext{%
  \footnotesize © 2025 IEEE. Personal use of this material is permitted. Permission from IEEE must be obtained for all other uses, in any current or future media, including reprinting/republishing this material for advertising or promotional purposes, creating new collective works, for resale or redistribution to servers or lists, or reuse of any copyrighted component of this work in other works. 
  This paper has been accepted for publication in {LASCAS 2025: 16th IEEE CASS Latin American Symposium on Circuits and Systems}.
  }
  
\newcommand\copyrightnotice{%
\begin{tikzpicture}[remember picture,overlay]
\node[anchor=south,yshift=10pt] at (current page.south) {\fbox{\parbox{\dimexpr\textwidth-\fboxsep-\fboxrule\relax}{\copyrighttext}}};
\end{tikzpicture}%
}

\maketitle

\begin{abstract}
Multi-Party Computation in the Head (MPCitH) algorithms are appealing candidates in the additional US NIST standardization rounds for Post-Quantum Cryptography (PQC) with respect to key sizes and mathematical hardness assumptions.
However, their complexity presents a significant challenge for platforms with limited computational capabilities. 
To address this issue, we present, to the best of our knowledge, the first design space exploration of MiRitH, a promising MPCitH algorithm, for embedded devices.
We develop a library of mixed HW/SW blocks on the Xilinx ZYNQ 7000, and, based on this library, we explore optimal solutions under runtime or FPGA resource constraints for a given public key infrastructure.
Our results show that MiRitH is a viable algorithm for embedded devices in terms of runtime and FPGA resource requirements.
\end{abstract}

\begin{IEEEkeywords}
PQC, MPCitH, Cryptography, FPGA
\end{IEEEkeywords}

\copyrightnotice

\section{Introduction}
Quantum computers are expected to break state-of-the-art (SoA) public key cryptosystems (PKCs) within the next decade~\cite{piani2023quantum}.
While this may seem far off in the future, the preparation of cryptographic infrastructures must begin today to account for mitigation time, development time, and product lifecycles (e.g., in the automotive industry).
For this reason, the US National Institute of Standards and Technology (NIST) is carrying out a Post-Quantum Cryptography (PQC) standardization process consisting of multiple rounds, during which four candidates were already selected for standardization.
However, due to their novelty, the ability to seamlessly replace them if they turn out to be insecure (crypto-agility) is very important.
Therefore, NIST started an additional round for PQC that relies on disjoint mathematical assumptions~\cite{nist2023}.

In this additional round, Multi-Party Computation in the Head (MPCitH) based digital signature algorithms (DSAs), such as MinRank in the Head (MiRitH)~\cite{mirith2023}, are promising candidates in terms of security, but they have higher computational complexity than the four selected algorithms.
This complexity is challenging for embedded devices as (i) they have limited computational power, (ii) multiple cryptographic algorithms may need to be supported (crypto-agility), (iii) cryptography is generally not the primary objective, and (iv) they often have tight application constraints with respect to runtime or hardware resources.

The efficient implementation of PQC algorithms has already been subject to various papers.
In~\cite{karl2024post}, a hardware/software (HW/SW) co-design approach was used to improve the implementation efficiency of the selected algorithms Dilithium and \textsc{Falcon} on a custom RISC-V platform, whereas in~\cite{schmid2023falcon} the first standalone HW accelerator for \textsc{Falcon} was proposed.
In~\cite{amiet2020fpga}, SPHINCS+, the selected hash-based algorithm, was implemented on HW.
From the additional rounds, only one code-based MPCitH algorithm - Syndrome Decoding in the Head (SDitH) - has been investigated for HW efficiency so far~\cite{deshpande2024sdith}.
Furthermore, in~\cite{kannwischer2024pqm4}, PQC algorithms from all rounds were benchmarked on an ARM Cortex M4 processor, and the results show that MiRitH is more than 41$\times$ slower than Dilithium.
To address these issues, we present in this paper, to the best of our knowledge, the first design space exploration of MiRitH on the Xilinx Zynq 7000 System on Chip (SoC).
In summary, the novel contributions of this work are:
\begin{enumerate}
    \item
    We perform an in-depth investigation of different HW/SW partitionings with respect to runtime and resource constraints.
    \item
    We formulate two optimization problems that reflect the previously described requirements (i)--(iv) and develop a library of HW/SW blocks for MiRitH to provide solutions to these problems.
    \item We compare the implementation results of our new methodology to SoA HW implementations of other PQC signature schemes.
\end{enumerate}

\section{Background}
DSAs are PKCs that consists of three functions, Key-Generation (\texttt{KeyGen}), Signing (\texttt{Sign}), and Verifying (\texttt{Open}).
MiRitH's security is based on the hardness of solving a MinRank problem~\cite{mirith2023}:
Given a $(k+1)$-tuple of matrices $\textbf{M}=(\textbf{M}_0, \textbf{M}_1, ..., \textbf{M}_k) \in (\mathbb{F}^{m \times n}_{q})^{k+1}$, where $\mathbb{F}_q$ denotes a finite field, and $\textbf{M}_i=[\textbf{M}_i^L|\textbf{M}_i^R]$ with the left-side matrix $\textbf{M}_i^L \in \mathbb{F}^{m \times (n-r)}_{q}$ and the right-side matrix $\textbf{M}_i^R \in \mathbb{F}^{m \times r}_{q}$, find $\alpha = (\alpha_1, ..., \alpha_k) \in \mathbb{F}_q^k$ and a matrix $\textbf{K} \in \mathbb{F}_q^{r \times (n-r)}$ such that:
\begin{equation}
    \label{eq:matrix_mul_triple}
    \textbf{M}_0^L + \sum_{i=1}^k \alpha_i \textbf{M}_i^L = (\textbf{M}_0^R + \sum_{i=1}^k \alpha_i \textbf{M}_i^R) \cdot \textbf{K}.
\end{equation}

MiRitH employs the Multi-Party Computation (MPC) protocol, where $N$ mutually distrusting parties prove the possession of their private views.
However, instead of involving multiple external parties, in MPCitH a single signer and a single verifier simulate the interaction of $N$ parties over $\tau$ rounds.
In short, MiRitH employs the following phases~\cite{mirith2023}:
\begin{itemize}
    \item \textbf{Phase 1 - Input preparation:} For each $l \in \{1,...,\tau\}, i \in \{1,...,N\}$, the signer sets up the private views using pseudo-random number generators (PRNGs).
    The views represent additive shares $\llbracket \alpha^{(l)} \rrbracket_i$ s.t. $\alpha^{(l)} = \sum_{i=1}^N \llbracket \alpha^{(l)} \rrbracket_i$.
    \item \textbf{Phase 2 - First challenge:} The first challenge is calculated by hashing the message to be signed, a salt and random shares from Phase 1.
    \item \textbf{Phase 3 - MPC execution:} The signer simulates the MPC protocols with the initial inputs and the first challenge by calculating solutions to the MinRank problem.
    \item \textbf{Phase 4 - Second challenges:} The signer computes a second challenge for each round $i^* \in \{1,...,N\}$ that corresponds to a randomly chosen party. 
    \item \textbf{Phase 5 - Signature:}
    $i^*$ determines which shares are part of the assembled signature and which are kept secret for security proofing by the verifier. 
\end{itemize}

\section{Design Rational}
Most SoA implementations of PQC focus on optimizing the runtime of the three DSA functions or on achieving advancements in some quality metric, e.g., FPGA slices$\times$delay~\cite{deshpande2024sdith}.
In this work, we take a different approach.
In real-world applications, the public key infrastructure (PKI) determines how often each peer performs the \texttt{Sign} and \texttt{Open} per connection.
In a typical Transport Layer Security (TLS) based PKI, the server provides a certificate chain, requiring the client to verify multiple signatures.
The client only needs to prove its identity when mutual authentication is demanded, which then requires only one \texttt{Sign} operation.
Furthermore, depending on the PKI, each peer may use a different algorithm for signing, and MiRitH may only be required for either \texttt{Sign} or \texttt{Open}.
We account for this fact by considering the total runtime $T$ with the factors $n$ and $m$:
\begin{equation}
    T = n \cdot t_{sign}+m \cdot t_{open},~n\in[0,1],~m \in \mathbb{N}_0.
\end{equation}

Embedded applications are often time or resource constrained, thus, the two explored optimization problems are:
\begin{enumerate}
    \item \textbf{Resource constraint:} Typically, acceleration is required for more than just MiRitH, e.g., to facilitate the implementation of multiple PQC algorithms (crypto-agility).
    In this case, a strictly constrained amount of FPGA resources $R_c$ is available.
    Using $R$$<$$R_c$ resources provides no benefit in this case, as the remaining ones will be left unused, and the goal is to minimize $T$ under $R_c$:
\begin{align}
    \mathcal{P}_R&: \min(n \cdot t_{sign}+m \cdot t_{open})~~ \text{s.t.}~~R \leq R_c.
\end{align}
    $R$ denotes all restricted FPGA resources: Lookup-Tables (LUTs), Flip-Flops (FFs), Block Random Access Memories (BRAMs) and Digital Signal Processor (DSP) slices.
    \item \textbf{Time constraint:} 
    Analogous to $\mathcal{P}_R$, the application may have strict constraints on the runtime $T_c$ of MiRitH, but there is no benefit if $T<T_c$, e.g., because the impact of the network delay dominates beyond $T_c$~\cite{schoffel2021energy}:
    \begin{align}
        \mathcal{P}_T&: \min(R)~~ \text{s.t.}~~(n \cdot t_{sign}+m \cdot t_{open}) \leq T_c.
    \end{align}
\end{enumerate}

In the following, solutions to $\mathcal{P}_T$ and $\mathcal{P}_R$ are developed through a library of mixed HW/SW implementations.

\section{HW/SW Co-Design}
We employed the following methodology:
\begin{enumerate}[A.]
    \item Profile the MiRitH \texttt{KeyGen}, \texttt{Sign} and \texttt{Open} functions in software to determine the computational bottlenecks.
    \item Determine promising HW/SW partitionings based on the profiling results and on a data dependency analysis.
    \item Design and implement accelerators for the partitionings on the Xilinx ZYNQ 7000.
\end{enumerate}
The steps have been executed iteratively to reach the final results presented and are discussed in the following.

\subsection{Profiling Results}
Table~\ref{tab:profiling} presents the profiling results of the C reference SW as provided in~\cite{mirith2023}.
We compiled the SW with $-O3$ and measured 10k runs for each DSA function.
The values show the time spent within each function excluding their subfunctions. 
As demonstrated, the major contributors to the total execution time are the operations on matrices and the \textsc{Keccak} algorithm.
In the matrix arithmetic, the largest computational bottleneck is the sum of scalar-matrix products $ \llbracket \textbf{E}^{(l)}\rrbracket_i=\sum_{j=1}^k \llbracket \alpha_j^{(l)} \rrbracket_i \textbf{M}_j$, see Eq. (1), which is denoted as $ \sum \alpha\textbf{M}$ in the following.
The second major contributor is \textsc{Keccak}, which is employed as a hashing algorithm and as a PRNG, e.g., to create random shares in Phase 1.

\begin{table}
    \centering
    \caption{Profiling results in ms and as a percentage of the total computation time of the respective DSA function.}
    \begin{tabular}{llll}
         & \textbf{\texttt{KeyGen}} & \textbf{\texttt{Sign}} & \textbf{\texttt{Open}} \\
         & [ms] &  [ms] & [ms] \\
         \toprule
         Mat. Arith. & 0.29 (16.22\%)
& 193.71 (79.45\%)
& 182.68 (81.24\%)
\\
         $^{\llcorner}$$ \sum \alpha \textbf{M}$ & $^{\llcorner}$0.27 (15.10\%)& $^{\llcorner}$\textbf{175.06} (\textbf{71.80\%})& $^{\llcorner}$\textbf{167.3} (\textbf{74.40}\%)\\
         $^{\llcorner}$Mat. Prod. & $^{\llcorner}$0.02 (1.12\%)& $^{\llcorner}$18.65 (7.65\%)& $^{\llcorner}$15.38 (6.84\%)\\
         \textsc{Keccak} & 1.48 (82.76\%)
& 46.94 (19.25\%)
& 39.24 (17.45\%)
\\
         $^{\llcorner}$Permute & $^{\llcorner}$\textbf{1.22} (\textbf{68.00\%})& $^{\llcorner}$40.23 (16.50\%)& $^{\llcorner}$32.61 (14.50\%)\\
         $^{\llcorner}$Squeeze & $^{\llcorner}$0.26 (14.60\%)& $^{\llcorner}$2.68 (1.10\%)& $^{\llcorner}$3.22 (1.43\%)\\
         $^{\llcorner}$Absorb  & $^{\llcorner}$0 (0.16\%)& $^{\llcorner}$4.02 (1.65\%)& $^{\llcorner}$3.42 (1.52\%)\\
         Others                 & 0.02 (1.03\%)
& 3.17 (1.30\%)& 2.95 (1.31\%)\\
        \midrule
         \textbf{Total}                  & 1.79 (100.00\%)& 243.82 (100.00\%)& 224.87 
         (100.00\%)\\
         \bottomrule
    \end{tabular}
    \label{tab:profiling}
\end{table}

\subsection{HW/SW Partitioning}

\begin{figure}
	\centering
		\includegraphics[width=0.5\textwidth]{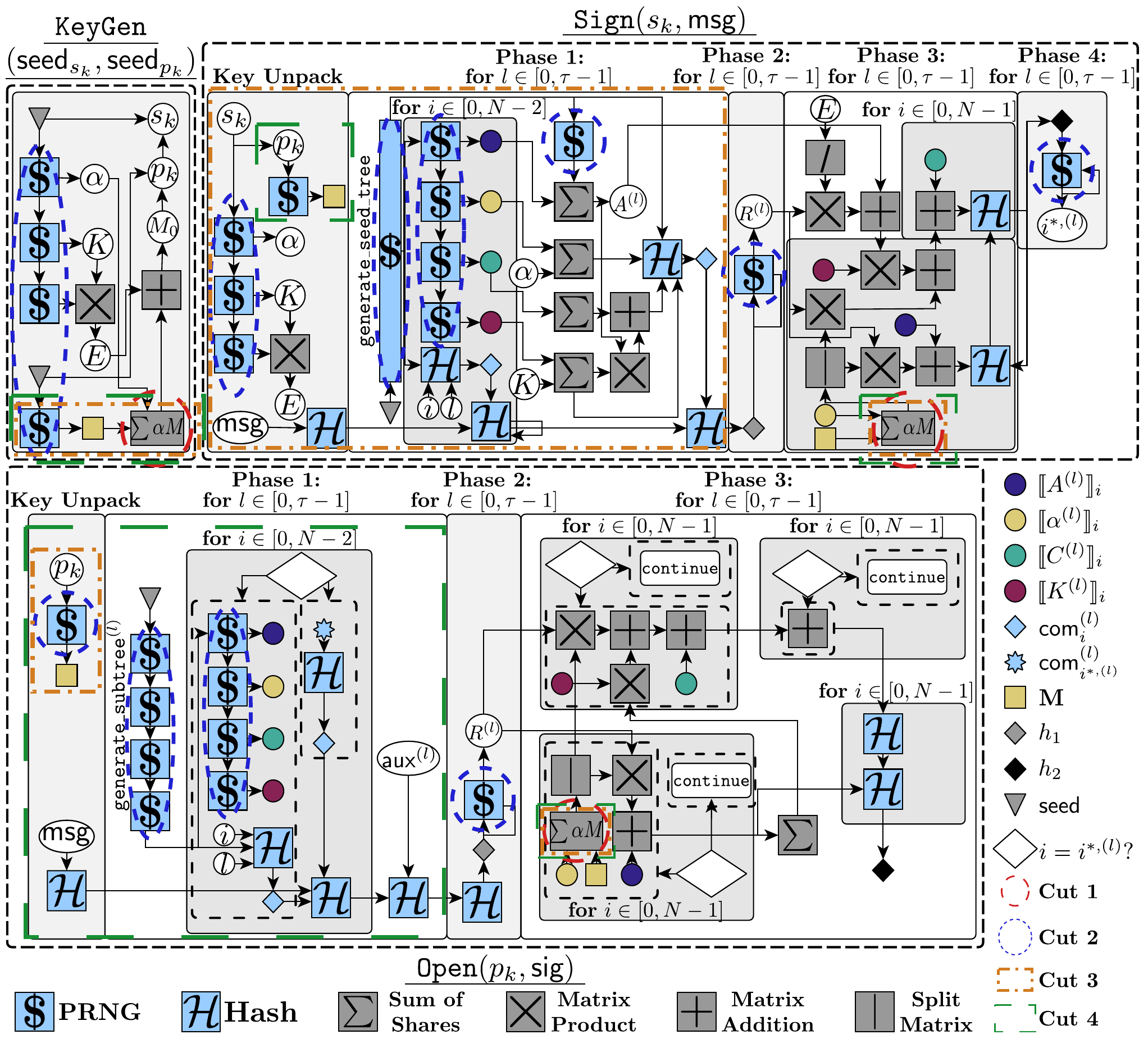}
		\caption{Simplified sketch of the datapath in the \texttt{KeyGen}, \texttt{Sign} and \texttt{Open} operation of MiRitH.}
	\label{fig:datapath}
\end{figure}

Fig.~\ref{fig:datapath} shows the datapath of the three DSA algorithms.
The blue-colored boxes correspond to functions related to \textsc{Keccak}, while the gray-colored boxes refer to matrix operations.
The partitions ("Cuts") between the HW and SW are represented by the different colored contours.
For clarity, the figure shows only four of the eight cuts examined, while Table~\ref{tab:cuts} shows all the cuts examined.
As $\sum \alpha \textbf{M}$ is the primary contributor to \texttt{Sign} and \texttt{Open}, Cut~1 is a crucial candidate to achieve significant speedups through HW acceleration.
Fig.~\ref{fig:datapath} shows that $\llbracket \alpha^{(l)} \rrbracket_i$ and $\textbf{M}$ are generated during the key unpacking and Phase 1, while the results of $\sum \alpha \textbf{M}$ are only required in Phase 3, therefore allowing for a concurrent execution of the Phase~1 iterations in SW and $\sum \alpha \textbf{M}$ in HW.
The objective of Cuts 3 and 6 is to reduce the runtime of the \texttt{Sign} operation while still providing substantial speed-ups for \texttt{Open} and \texttt{KeyGen} through the use of $\sum \alpha \textbf{M}$ and the public key unpacking module.
Cuts 4 and 7 employ the same concept for the \texttt{Open} operation.
In contrast, Cuts 5 and 8 minimize the runtime of both \texttt{Sign} and \texttt{Open} operations, making them well-suited for PKIs where both are required.

\begin{table}
    \centering
    \caption{Partitions for MiRitH. Cut refers to the part of the algorithm executed on dedicated HW accelerators.}
    \begin{tabularx}{\linewidth}{l|X}
        \textbf{Partition} & \textbf{HW Accelerator}\\
        \hline
        Cut 1 & Sum of scalar-matrix products $\sum \alpha \textbf{M}$ \\
        Cut 2 & \textsc{Keccak}-based PRNG \\
        Cut 1+2 & $\sum \alpha \textbf{M}$ + PRNG \\
        Cut 3 & secret/public key unpack, sign Phase 1, $\sum \alpha \textbf{M}$\\
        Cut 4 & public key unpack, open Phase 1, $\sum \alpha \textbf{M}$ \\
        Cut 5 & secret/public key unpack, sign and open Phase 1, $\sum \alpha \textbf{M}$ \\
        Cut 6 & secret/public key unpack, sign Phase 1--4, $\sum \alpha \textbf{M}$ \\
        Cut 7 & public key unpack, open Phase 1--4, $\sum \alpha \textbf{M}$ \\
        Cut 8 & secret/public key unpack, sign and open Phase 1--4, $\sum \alpha \textbf{M}$ \\
    \end{tabularx}
    \label{tab:cuts}
\end{table}

\subsection{HW Design}

\begin{figure}[t]
\centering
     \includegraphics[width=0.4\textwidth]{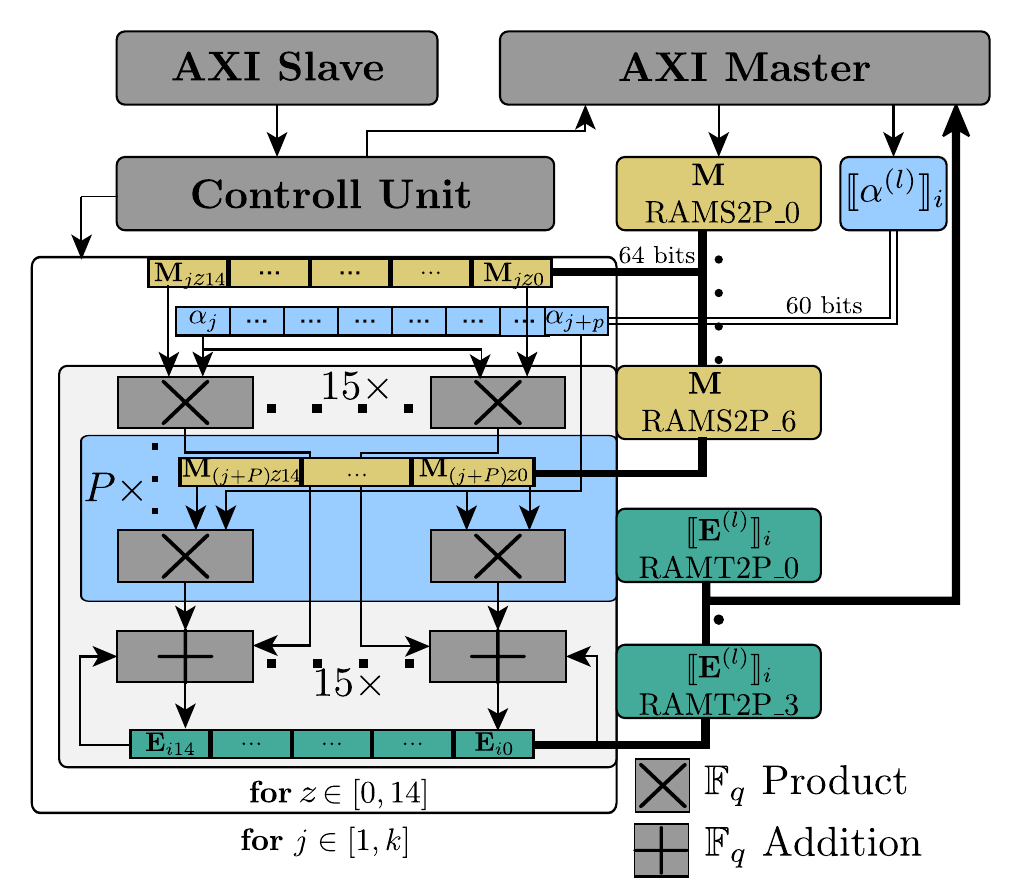}
    \caption{Block diagram of the $\textbf{E}=\sum \alpha \textbf{M}$ HW accelerator.}
\label{fig:matmul}
\end{figure}

Fig.~\ref{fig:matmul} shows the HW design of the $\sum \alpha \textbf{M}$ accelerator.
To enhance concurrency, we store the matrix $\textbf{M}$ in 60-bit words, thereby enabling the access to a complete matrix column comprising 15 4-bit finite field elements within a single memory location.
The vector of scalars $\alpha$ is stored similarly.
During each computation step, the result of one scalar-column multiplication is calculated and added to the intermediate result $\textbf{E}$.
The parallelization factor $P$, can be configured prior to the HW synthesis, and allows to process $P$ scalar-column products in parallel.
Regarding the PRNG HW module, its most important component is the \textsc{Keccak} Permute function, which is computed in 24 cycles.
The remaining cuts are combined into one modular HW design, where each of the required blocks can be selected prior to HW synthesis.

\section{Results}
We used high-level synthesis (HLS) with AMD Xilinx Vitis 2023.2 to synthesize the hardware.
All HW/SW blocks were validated against the C reference SW on a Xilinx ZYNQ xc7z020clg400-1. 
The reported runtimes are based on the average value of 10k tests for each DSA function.
The standard deviation was in most cases less than 1\textperthousand.
Table~\ref{tab:results} shows the available configurations in our developed MiRitH library.
The achieved target frequency was 142 MHz for the basic blocks (Cut 3, 4 and 5) and 125 MHz for the others.
DSPs are not included as at most~2 (1\%) were used.

\begin{table}
    \centering
    \caption{Results of the implementation of the cuts described in Table~\ref{tab:cuts}. $P$=1 if not specified. The FPGA resource percentages represent the utilization of the total available resources. The blue boxes highlight the optimized columns.}
    \label{tab:results}
    \begin{adjustbox}{max width=0.5\textwidth}
    \begin{tabular}{llrrrrr}
        \textbf{Partition} &  \textbf{\texttt{KeyGen}} & \textbf{\texttt{Sign}} & \textbf{\texttt{Open}} & \multicolumn{3}{c}{\textbf{FPGA Resources}} \\
    & [ms]& [ms] & [ms] & [kLUTs]& [kFFs]& [BRAMs] \\
    \toprule
    \multicolumn{7}{l}{\textbf{Basic Blocks}} \\
     SW only& 1.79& 243.82 &224.87& -& -&-
\\
     Cut 1& 1.19& 69.62& 61.85 & 3.8 (7\%) & 5.4 (5\%) &9 (6\%)
\\
     Cut 2& 1.05& 230.92& 218.28& 9 (17\%)
& 8.6 (8\%)
&6.5 (5\%) 
\\
     Cut 1+2& 0.88& 53.16& 53.35& 13.0 (24\%)& 14.5 (14\%)&15.5 (11\%)\\
     \midrule
     \multicolumn{7}{l}{\textbf{\texttt{Sign} Optimized}} \\
        Cut 3& 0.07& \cellcolor{lightgray}40.52& 60.67& 22.3 (42\%)
& 25.1 (24\%)
&15.5 (11\%)
\\
     Cut 6& 0.12& \cellcolor{lightgray}11.75& 60.51& 26.0 (49\%)
& 26.0 (24\%)
&59 (42\%)
\\
     Cut 6, $P$=4& 0.11& \cellcolor{lightgray}6.36& 60.59& 26.0 (49\%)
& 25.1 (24\%)
&61 (44\%)
\\
     Cut 6, $P$=8& 0.11& \cellcolor{lightgray}5.62& 60.57& 29.4 (55\%)
& 25.7 (24\%)&64 (46\%)
\\
     Cut 6, $P$=16& 0.11& \cellcolor{lightgray}5.23& 60.56& 34.0 (64\%)& 26.4 (25\%)&73 (52\%)\\
     \midrule 
     \multicolumn{7}{l}{\textbf{\texttt{Open} Optimized}} \\
          Cut 4& 0.07& 72.00& \cellcolor{lightgray}39.57& 19.3 (36\%)
& 22.4 (21\%)
&15 (11\%)
\\
     Cut 7& 0.10& 68.52& \cellcolor{lightgray}11.52& 22.2 (42\%)
& 23.8 (22\%)
&59.5 (43\%)
\\
     Cut 7, $P$=4& 0.11& 68.52& \cellcolor{lightgray}6.50& 22.2 (42\%)
& 23.1 (22\%)
&61 (44\%)
\\
     Cut 7, $P$=8& 0.11& 68.32& \cellcolor{lightgray}5.81& 25.6 (48\%)& 23.6 (22\%)&66 (47\%)
\\
     Cut 7, $P$=16& 0.10& 68.25& \cellcolor{lightgray}5.45& 26.6 (50\%)& 24.3 (23\%)&74 (53\%)\\
     \midrule
     \multicolumn{7}{l}{\textbf{\texttt{Sign} + \texttt{Open} Optimized}} \\
      Cut 5& 0.07& \cellcolor{lightgray}40.24& \cellcolor{lightgray}37.38& 25.4 (48\%)& 28.5 (27\%)&18 (13\%)
\\
     Cut 8& 0.12& \cellcolor{lightgray}11.75& \cellcolor{lightgray}11.62& 30.9 (58\%)& 33.0 (31\%)&63.5 (45\%)
\\
     Cut 8, $P$=4& 0.11 & \cellcolor{lightgray}6.36& \cellcolor{lightgray}6.56 & 31.7 (60\%)& 32.4 (30\%)&65 (46\%)
\\
     Cut 8, $P$=8& 0.11& \cellcolor{lightgray}5.62& \cellcolor{lightgray}5.87& 34.8 (65\%)& 32.4 (30\%)&69 (49\%)
\\
     Cut 8, $P$=16& 0.11& \cellcolor{lightgray}5.23&\cellcolor{lightgray}5.51& 35.6 (67\%)& 33.1 (31\%)&78 (56\%)\\
     \bottomrule
    \end{tabular}
    \end{adjustbox}
\end{table}

\subsection{Impact of different Cuts on Runtime of \texttt{Sign} and \texttt{Open}}
Cut 1 reduces the execution time by a factor of 3.5 (\texttt{Sign}) and 3.65 (\texttt{Open}) compared to the SW only implementation.
This reduction in execution time corresponds to the time spent on $\sum \alpha \textbf{M}$ determined in the profiling (Table~\ref{tab:profiling}), thus proving that we maximize HW/SW concurrency by moving its computation into Phase 1.
On the other hand, Cut 2 yields a minor improvement for both functions without Cut 1.
The largest speedups are achieved by combining all phases and the key unpack module (Cuts 6, 7, and 8) and by choosing $P$=$16$ for the $\sum \alpha \textbf{M}$ HW module.

\begin{figure}[t]
\centering
    \includegraphics[width=0.4\textwidth]{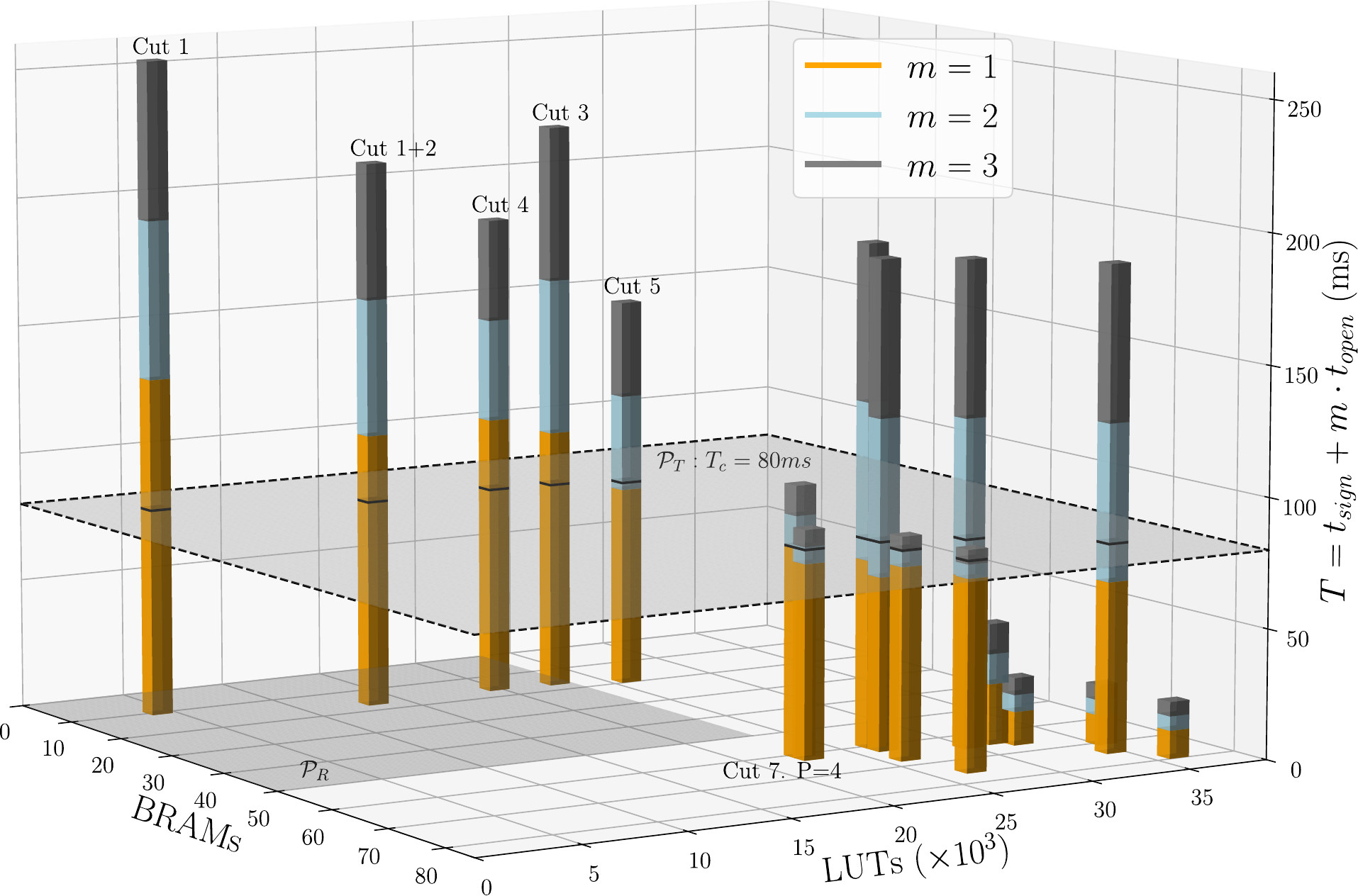}
    \caption{Visualization of optimization constraints. For $\mathcal{P}_T$, the intersection between the plane $T_c$=80~ms and the bars of each Cut is shown in black. The $\mathcal{P}_R$ constraints correspond to the highlighted grid area.}
\label{fig:opt}
\end{figure}

\subsection{Solutions for $\mathcal{P}_R$ and $\mathcal{P}_T$}
Due to the limited size of the library, exhaustive search is feasible to solve both optimization problems.
For example, suppose $\mathcal{P}_R$ with $R_c$=23k LUTs, 30k FFs, and 50 BRAMs, with $n,m$=1.
In this case, Cut 3 provides the optimum with $T$=101.19~ms.
However, if $n$=$1$,$m$=$2$, the optimal solution with the same $R_c$ is Cut~4 with $T$=151.14~ms versus $T$=161.86~ms for Cut~3.
This is shown in Fig.~\ref{fig:opt}.
Relaxing the LUT constraint to 26k allows to use Cut 5 with a significantly lower runtime of $T$=115.8~ms, and also relaxing the BRAM constraint to 70 allows to use Cut 7, $P$=8 with $T$=79.94~ms.

For $\mathcal{P}_T$, assume that $T_C$=110~ms due to the network delay, $n$,$m$=1, and the goal is to minimize the number of LUTs used.
While most solutions satisfy $T_C$=110~ms, Cut~1+2 is optimal (in this library) with respect to LUTs.
When $T_C$=80~ms, however, the best solution in terms of LUTs is Cut 7, $P$=4, while the best solution in terms of BRAMs that satisfies $T_C$ is Cut 5.
This is also visualized in Fig.~\ref{fig:opt}.
In summary, the results show that the optimal solution strongly depends on the assumed application constraints as well as on the PKI.

\begin{table}
    \centering
    \caption{Runtime and required FPGA resources of SoA Implementations.}
    \label{tab:soa}
    \begin{tabular}{lrrrll}
        \textbf{Impl.} & \textbf{\texttt{KeyGen}} & \textbf{\texttt{Sign}} & \textbf{\texttt{Open}} & \textbf{FPGA Resources} \\
        & [ms] & [ms] & [ms] & LUT/FF/BRAM/DSP \\
         \toprule
         Dil2~\cite{karl2024post} & 5.9 & 19.1 & 6.5 & 22k / 13k / 6 / 13 \\
         Fal512~\cite{karl2024post} & - & - & 3.1 & 22k / 13k / 6 / 13 \\
         Fal512~\cite{schmid2023falcon} & - & 4.2 & - & 47k / 44k / 56 / 182 \\
         Fal512~\cite{schmid2023falcon} & - & - & 0.6 & 11k / 8k / 13 / 15 \\
         SPH+128s~\cite{amiet2020fpga} & - & 12.4 & 0.7 &  48k / 72k / 12 / 0 \\
         SDitH-L1~\cite{deshpande2024sdith} & 0.3 & 41.0 & 8.6 & 17k / 9k / 165 / 0 \\
    \bottomrule
    \end{tabular}
\end{table}

\subsection{Comparison to State-of-the-Art}
Table~\ref{tab:soa} shows a selection of SoA implementations.
Compared to the implementations of \textsc{Falcon} and SPHINCS+128s, our implementation requires significantly less resources to achieve comparable or shorter runtimes for the \texttt{Sign} operation.
Compared to SDitH, our optimized designs achieve shorter runtimes (3$\times$ for \texttt{KeyGen}, 8$\times$ for \texttt{Sign} and 1.6$\times$ for \texttt{Open}) using Cut 8, requiring 80\% more LUTs and 3$\times$ the FFs, but 60\% less BRAMs.
 
Our results also indicate that the performance gap between Dilithium and MiRitH may be smaller in HW/SW co-design approaches than in the software benchmark from~\cite{kannwischer2024pqm4}, where MiRitH required 41$\times$ more time for \texttt{Sign} and 133$\times$ more time for \texttt{Open}.
The library presented in this paper allows for the execution of \texttt{Sign} to be up to 4$\times$ faster than Dilithium's \texttt{Sign} in the SoA HW/SW co-design~\cite{karl2024post}, requiring 30\% more LUTs, 10× more BRAMs, and 11 fewer DSPs.
If resource utilization is tightly restricted, Cut 1 already reduces the disparity by more than one order of magnitude (3$\times$ slower instead of 41$\times$ for \texttt{Sign} and 9$\times$ slower for \texttt{Open} instead of 133$\times$) with 83\% fewer LUTs and 59\% fewer FFs than~\cite{karl2024post}.
Also, except of~\cite{schmid2023falcon}, the results shown have been implemented directly using hardware description language (HDL), which is known to outperform HLS-generated code in terms of resource efficiency.
Moreover, the cited results employ more powerful FPGAs, enabling them to attain higher clock frequencies. 
Thus, a direct comparison with our solution is pessimistic.
 
\section{Conclusion}
In this work, we performed a design space exploration of MiRitH, a PQC DSA candidate in the additional round of the NIST PQC process in the context of embedded devices.
A library of HW/SW blocks for the Xilinx ZYNQ 7000 was presented to find the optimal solution, either in terms of minimum FPGA resource utilization or minimum runtime for a given PKI.
To the best of our knowledge, this approach is unprecedented in the context of PQC, and our paper represents the first HW implementation of a MinRank scheme of the additional round.
Our results prove that MiRitH is a promising candidate for embedded devices when HW implementations are considered.
Future work could extend this library to include more parameter sets of MiRitH, consider side-channel attacks and thus incorporate the security perspective into the library trade-offs, or further increase the implementation efficiency by using HDLs as opposed to our HLS-based approach.

\newpage
\bibliographystyle{IEEEtran}
\bibliography{main}

\begin{thebibliography}{1}
\providecommand{\url}[1]{#1}
\csname url@samestyle\endcsname
\providecommand{\newblock}{\relax}
\providecommand{\bibinfo}[2]{#2}
\providecommand{\BIBentrySTDinterwordspacing}{\spaceskip=0pt\relax}
\providecommand{\BIBentryALTinterwordstretchfactor}{4}
\providecommand{\BIBentryALTinterwordspacing}{\spaceskip=\fontdimen2\font plus
\BIBentryALTinterwordstretchfactor\fontdimen3\font minus \fontdimen4\font\relax}
\providecommand{\BIBforeignlanguage}[2]{{%
\expandafter\ifx\csname l@#1\endcsname\relax
\typeout{** WARNING: IEEEtran.bst: No hyphenation pattern has been}%
\typeout{** loaded for the language `#1'. Using the pattern for}%
\typeout{** the default language instead.}%
\else
\language=\csname l@#1\endcsname
\fi
#2}}
\providecommand{\BIBdecl}{\relax}
\BIBdecl

\bibitem{piani2023quantum}
M.~Mosca and P.~Marco, ``{Quantum Threat Timeline Report 2023},'' 2023.

\bibitem{nist2023}
{US National Institute of Standards and Technology (NIST)}, ``{Call for Additional Digital Signature Schemes for the Post-Quantum Cryptography Standardization Process},'' \url{https://csrc.nist.gov/csrc/media/Projects/pqc-dig-sig/documents/call-for-proposals-dig-sig-sept-2022.pdf}, 2022, Retrieved 2024-03-13.

\bibitem{mirith2023}
{Gora Adj, Luis Rivera-Zamarripa, Javier Verbel}, ``{MiRitH (MinRank in the Head)},'' {2023}, \url{https://pqc-mirith.org/assets/downloads/mirith_specifications_v1.0.0.pdf}, Retrieved 2024-08-08.

\bibitem{karl2024post}
P.~Karl, J.~Schupp, T.~Fritzmann, and G.~Sigl, ``{Post-Quantum Signatures on RISC-V with Hardware Acceleration},'' \emph{ACM Transactions on Embedded Computing Systems}, vol.~23, no.~2, pp. 1--23, 2024.

\bibitem{schmid2023falcon}
M.~Schmid, D.~Amiet, J.~Wendler, P.~Zbinden, and T.~Wei, ``{Falcon Takes Off-A Hardware Implementation of the Falcon Signature Scheme},'' \emph{Cryptology ePrint Archive}, 2023.

\bibitem{amiet2020fpga}
D.~Amiet, L.~Leuenberger, A.~Curiger, and P.~Zbinden, ``{Fpga-based Sphincs+ Implementations: Mind the Glitch},'' in \emph{2020 23rd Euromicro Conference on Digital System Design (DSD)}.\hskip 1em plus 0.5em minus 0.4em\relax IEEE, 2020, pp. 229--237.

\bibitem{deshpande2024sdith}
S.~Deshpande, J.~Howe, J.~Szefer, and D.~Yue, ``{SDitH in Hardware},'' \emph{Cryptology ePrint Archive}, 2024.

\bibitem{kannwischer2024pqm4}
M.~J. Kannwischer, M.~Krausz, R.~Petri, and S.-Y. Yang, ``{pqm4: Benchmarking NIST Additional Post-Quantum Signature Schemes on Microcontrollers},'' \emph{Cryptology ePrint Archive}, 2024.

\bibitem{schoffel2021energy}
M.~Sch{\"o}ffel, F.~Lauer, C.~C. Rheinl{\"a}nder, and N.~Wehn, ``{On the Energy Costs of Post-Quantum KEMs in TLS-based low-power secure IoT},'' in \emph{Proceedings of the International Conference on Internet-of-Things Design and Implementation}, 2021, pp. 158--168.

\end{thebibliography}

\end{document}